\documentclass{acmsiggraph}
\title{Computed Axial Lithography (CAL):\\ Toward Single Step 3D Printing of Arbitrary Geometries}

\author{Brett E. Kelly$^{1,3}$, Indrasen Bhattacharya$^{2}$, Maxim Shusteff$^{3}$, Robert M. Panas$^{3}$, Hayden K. Taylor$^{1}$, and Christopher M. Spadaccini$^{3}$\\$^1$Department of Mechanical Engineering, University of California, Berkeley\\ $^2$Department of Applied Science and Technology, University of California, Berkeley\\$^3$Materials Engineering Division, Lawrence Livermore National Laboratory}
\pdfauthor{Brett Kelly, Indrasen Bhattacharya}


\begin{document}

%%% This is the ``teaser'' command, which puts an figure, centered, below 
%%% the title and author information, and above the body of the content.

\teaser{
   \includegraphics[height=2.0in]{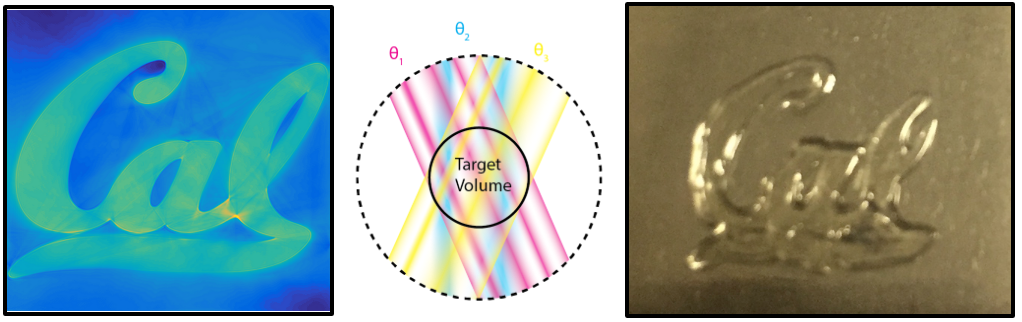}
   \caption{ Left: Optimized dose distribution for the Cal logo target, Middle: Schematic of the projection process, Right: Printed result with an optimized recipe}
 }

\maketitle

\begin{abstract}

Most additive manufacturing processes today operate by printing voxels (3D pixels) serially point-by-point to build up a 3D part. In some more recently-developed techniques, for example optical printing methods such as projection stereolithography \cite{PuSL}, \cite{Carbon3D}, parts are printed layer-by-layer by curing full 2d (very thin in one dimension) layers of the 3d part in each print step. There does not yet exist a technique which is able to print arbitrarily-defined 3D geometries in a single print step. If such a technique existed, it could be used to expand the range of printable geometries in additive manufacturing and relax constraints on factors such as overhangs in topology optimization. It could also vastly increase print speed for 3D parts. In this work, we develop the principles for an approach for single exposure 3D printing of arbitrarily defined geometries. The approach, termed Computed Axial Lithgography (CAL), is based on tomographic reconstruction, with mathematical optimization to generate a set of projections to optically define an arbitrary dose distribution within a target volume. We demonstrate the potential ability of the technique to print 3D parts using a prototype CAL system based on sequential illumination from many angles.  We also propose new hardware designs which will help us to realize true single-shot arbitrary-geometry 3D CAL.

\end{abstract}


%\ccsdesc[500]{Computing methodologies~Image manipulation}
%\ccsdesc[300]{Computing methodologies~Computational photography}


% The next three commands are required, and insert the user-generated keywords, 
% The CCS concepts list, and the rights management text.
% Please make sure there is a blank line between each of these three commands.

\keywordlist

%\conceptlist

%\printcopyright

\section{Introduction}

\subsection{Existing methods}

3D printing technology has expanded greatly since its inception in the 1980s. Today, additive fabrication techniques are used in a variety of fields to fabricate structures that often could not be manufactured by conventional subtractive processes. Applications range from biological scaffolds for culturing tissue in vitro to microarchitected materials with microstructures designed to produce enhanced or novel material properties \cite{PuSL}. Most 3D printing methods today operate in a similar form where a small quantum (voxel) of the final 3D part is generated in a single operation. This operation may involve for example extrusion through a nozzle (as in methods based on inkjetting or fused deposition modeling) or formation through an interaction with light (as in methods using laser sintering or photocrosslinking). In these technologies, a 3D part is typically formed by repetition of the unit print operation serially point-by-point throughout the user-defined 3D geometry. The unit print operation can thus be thought of as zero dimensional as it prints a single point (small 3D voxel) of the 3D volume. More recently, higher throughput techniques based on projection stereolithography have emerged \cite{PuSL},\cite{Carbon3D}. These methods show greatly increased print speed by printing a cross-section of material in a single print operation. This is accomplished using projection optics to generate a two-dimensional image which is then focused at the surface of a photocurable resin. The unit operation can then be thought of as two-dimensional (with a small finite thickness). The operation must again be repeated serially to build up a 3D part from thin parallel cross-sections. Following this trend of increased throughput, the next logical advance would be to develop a printing method where the unit operation is three-dimensional. There have been some successful attempts toward this end. However, the space of printable geometries for the 3D unit operation is severely limited to periodic structures with one of the three dimensions much smaller than the others, formed through interference lithography \cite{3DLith} or more recently structures formed from the superposition of 3 intersecting images \cite{SFF} . To date, to the best knowledge of the authors, there does not exist a 3D printing method which can print arbitrary 3D volumes with a single unit operation. This work lays the computational imaging foundations and provides experimental validation for a technique which could achieve single shot 3D stereolithography of almost arbitrary geometries.

\subsection{Advantages of single shot lithography}

It should be somewhat clear that single shot 3D lithography of arbitrary geometries could achieve greatly increased print speed compared to existing methods which rely on a zero or two-dimensional unit print operation. However, while an important motivating factor, speed is not the only potential advantage of such a technique. There still exist geometries which are difficult or fundamentally impossible to print using current additive manufacturing capabilities. The most clear case of this is in geometries with overhanging structures which cannot be printed without the use of a sacrificial support material. The field of topology optimization has blossomed with the advent of additive manufacturing technologies since these techniques have vastly expanded the space of printable geometries compared to subtractive cutting or machining techniques. However, there still exist manufacturing constraints, such as the overhang problem \cite{TopOpt}, which could potentially be overcome with the successful invention of a single shot stereolithography technique.

\subsection{Approach}
The approach used in this work is founded on the generation of an accumulated 3D intensity distribution designed carefully in conjuntion with photocuring chemistry such that in a single unit operation with a single development step, a 3D part of arbitrary user-defined geometry is generated. The approach demonstrated in this work, Computed Axial Lithography (CAL), is rooted in the concept of computed tomographic (CT) imaging and its application to stereolithography. In this work, prototype CAL systems are constructed which illuminate a photcurable resin sequentially from many anlges about a single axis. One CAL embodiment,  which captures the algorithmic essence of the technique is demonstrated using time multiplexed angular projection of 1D images into a 2D target volume. A second embodiment uses using time multiplexing of 2D images into a 3D volume and shows the ability to print 3D parts. A future CAL embodiment may use a microlens array with LED subpixels for simultaneous spatially and angularly resolved projection into the target volume. This will lead to single shot 3D volume printing.

\section{Image computation}

Rather than building up the target volume layer by layer (and voxel by voxel), this method proposes to expose the transparent resin by projecting 2D images from a grid of azimuthal angles. The method is based on one of the possible reconstruction procedures in computed tomography (CT). Prior art in a similar technique exists in intensity modulated radiation therapy (IMRT) for cancer treatment. IMRT seeks to produce a 3D variation of radiation dose in a target volume in the patient's body. Typical practice in this radiotherapy procedure involves the projection of a small number(7-10) of 2D distributions of X-ray pencil beams at a few selected angles. The intensity of the beamlets is chosen to satisfy a set of therapeutic constraints including: sufficient but not excessive dosage in the tumour, low radiation dose in certain critical organs as well as the physical constraint that radiation dose must be positive. Optimization techniques have yielded success in defining previously challenging dosage distributions, including concave regions such  as the prostate gland. Producing dose distributions in a clinically reasonable timeframe while satisfying multiple objectives continues to be an area of active research. The 3D printing problem has similarities in terms of constraints, but greater freedom in the exact dose distribution because of the thresholding behaviour of the resin. We would also like to draw attention to the wide applicability of some of the underlying concepts related to the Fourier slice theorem in areas as diverse as radio astronomy and lightfield photography. 

\subsection{Computed Tomography Reconstruction}
In the following discussion, we focus on the case of generating dose distributions in 2D flatland from 1D projections. We later extend this into the third dimension by concatenating calculated 1D projections for each Z-slice. We emphasize that all the beams for a particular projection angle are parallel in this raytracing picture. This is known as parallel beam tomography in the medical community. In the CT imaging configuration where a uniform pencil beam at azimuthal angle $\theta$ is projected into the 2D imaging volume with optical density given by $R(x,y)$, the collected dose distribution on the 1D camera along the space dimension x is given by $P(x,\theta)$:

\begin{equation}\label{Radon}
P(x,\theta) = \int R(x\cos(\theta)-u\sin(\theta),x\sin(\theta)+u\cos(\theta))du
\end {equation}

where $(-u\sin(\theta),u\cos(\theta))$ for parameter $u$ represents a line through the origin in the direction of exposure for the particular angle $\theta$. $P(x,\theta)$ is the well known Radon transform of the 2D image $R$. We will refer to $P(x,\theta)$ as the angular projections of the image. From the projection slice theorem, it turns out that the fourier transform of a projection at angle $\theta$ is exactly equal to a 1D sample of the original image's 2D fourier transform $R(k_x,k_y)$. The 1D sample is taken along a slice $(k\cos(\theta),k\sin(\theta))$ corresponding to the particular angle at which the projection was integrated.  This can be summarized as:

\begin{equation}\label{Radon}
\int P(x,\theta) e^{-ikx} dx = R(k\cos(\theta),k\sin(\theta))
\end {equation}

This is illustrated in Fig. 1, where the slices in the Radon transform on the left panel can be compared to similarly shaded central slices in the Fourier domain. 1D cuts along the radon transform and the corresponding Fourier slices have also been illustrated. We also note that sufficiently dense sampling in the Fourier domain is required for an accurate CT image reconstruction. Given $N$ radial samples in the Fourier domain, one possible conservative heuristic is to ensure that the maximum distance between slices is equal to the separation of sample points in the radial direction. This leads to $\pi N/2$ angular samples within 180 degrees of angular range. This is greater than the number of spatial samples, contrary to the convention in radiation therapy. In the following work and preliminary embodiment, we have used a large number of angular samples - typically twice or more than the number of spatial samples. In case the implementation favours fewer angular samples, we could optimize for a reconstruction using coarser angular sampling, or fewer angles.

\begin{figure}[ht]
  \centering
  \includegraphics[width=2.7in]{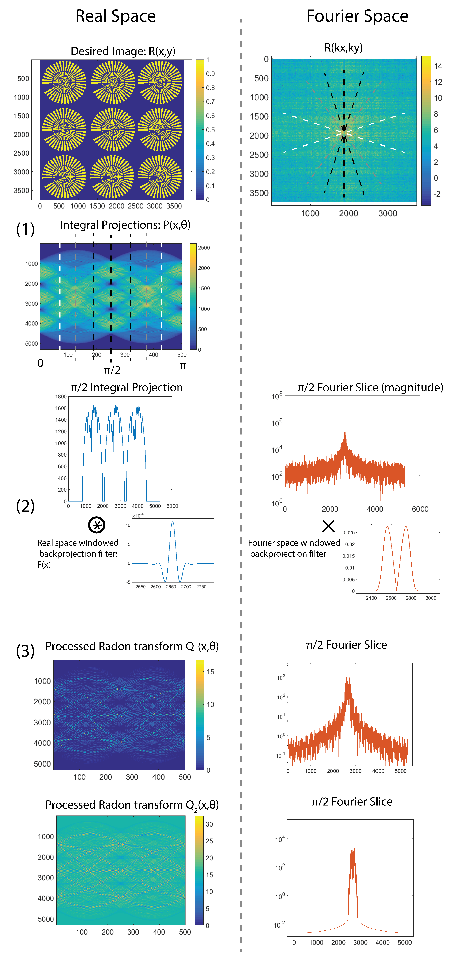}
  \caption{CT reconstruction example: top panel - the target 2D image is shown on left, with Fourier domain on right. (1) Particular angular projections correspond to central slices in the Fourier domain. (2) Backprojection filtering. (3) top: positivity enforced by setting negatives to zero, or bottom: adding offset.}
  \label{FourierProcess}
\end{figure}

The reconstruction of the 2D image volume follows from an algorithmic time reversal of the CT imaging process. This backprojection algorithm starts with the measured 1D projections and propagates each backwards while uniformly \textit{exposing} the target region with this intensity pattern. This is repeated for every angle. From the central slice theorem, this corresponds to building up the sample slice by slice in the Fourier domain. This algorithmic back-projection then motivates a technique to physically backproject the computed radon transform at each angle and directly construct desired 3D dose volumes. However, we need to invert the Jacobian required to transform the radial coordinate representation of the Fourier domain image to Cartesian coordinates. Equivalently, this can be considered as applying a radially increasing ramp filter in order to compensate for the inverse radial oversampling inherent in the Fourier slicing approach. Further, we also apply a window to the ramp filter in the Fourier domain, so as to exclude high spatial frequencies beyond the degree of sampling provided by the number of angular samples. IMRT literature suggests an exponential windowing filter for a smooth backprojection filter. This leads to the following \textit{backprojection filter} in the Fourier domain $(k,\theta)$:

\begin{equation}\label{Radon}
H(k) = |k|e^{-(\frac{k}{k_0})^4}
\end {equation}

where $k_0$ is chosen based on the number of angular samples [Ref]. This windowed high pass filter is applied on every Fourier slice before back-propagating it to form the image. The spatial domain representation of the filter has negative ripples that often lead to the backprojections being negative even if the projections themselves are positive. This is a major challenge since the backprojections are physically constrained to positive values. We have attempted to address this using a few different methods.

\begin{figure}[ht]
  \centering
  \includegraphics[width=3.3in]{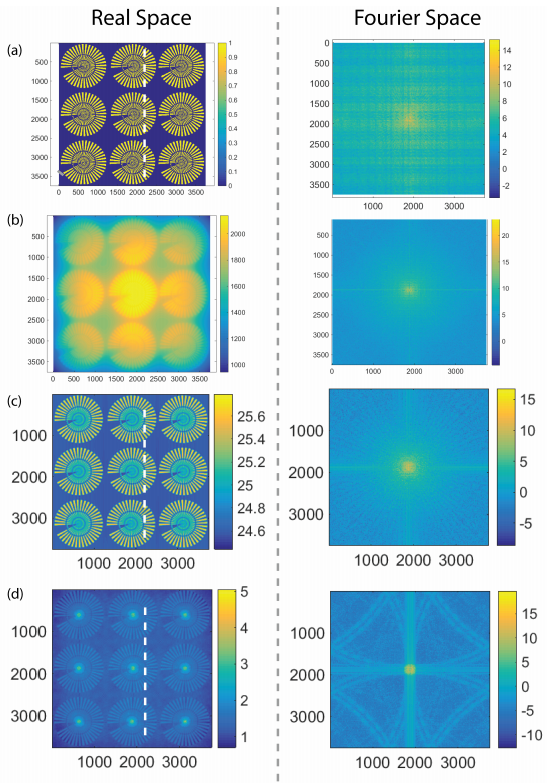}
  \caption{(a) Target image and fourier representation (in log-intensity scale), (b) Unfiltered backprojection showing emphacized low frequencies, (c) Zero offset shows good image shape, but with impractically low contrast, (d) Setting negatives to zero produces an image with more contrast, but sometimes arbitrary features}
  \label{Heuristics}
\end{figure}

As one preliminary heuristic method, we attempt to preserve the relative shape of the filtered backprojections $R_f(x,\theta)$ by adding $min(R_f(x,\theta))$ to the entire set of backprojections and performing an unfiltered inverse radon transform of this set of positive values so as to obtain the 2D dose distribution. However, as seen in Fig. 2(c) and the 1-D cuts in Fig. 3, this leads to the dose contrast being very low. We need a sufficient dose contrast between the cured and uncured regions so that the resin development process can operate. Another heuristic is to simply set the negative values in the backprojections to zero. This approach does lead to better contrast, however at the expense of accurate reproduction of the desired intensity pattern. The image cuts in Fig. 3 provide a specific example for the case of the Siemens star target. Although the heuristic results are not satisfactory, they do provide some intution, and possibly an initialization step for a constrained optimization procedure.

\begin{figure}[ht]
  \centering
  \includegraphics[width=3.3in]{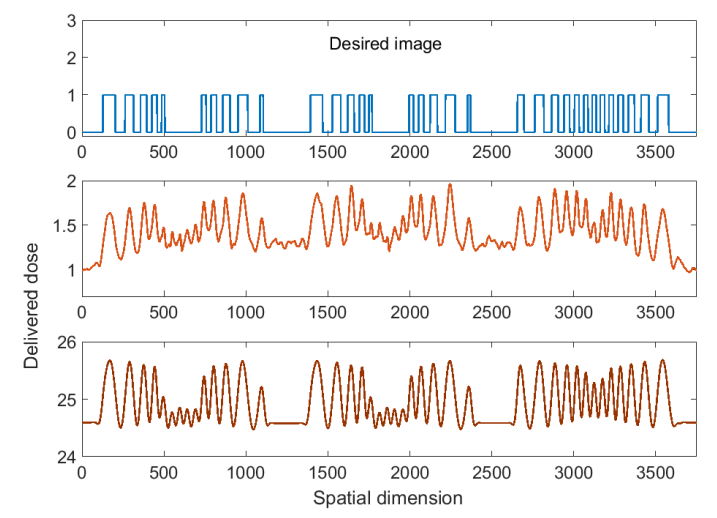}
  \caption{1D cuts from Siemens star example (along dashed white lines in Fig. 2): top - desired image, mid - setting negatives to zero, bottom - adding offset}
  \label{Heuristics_cuts}
\end{figure}

\subsection{Optimization Heuristic for Image Generation}

In order to frame the problem suitably for mathematical optimization,  we need a good understanding of the forward model that transforms 1D images at each angle into a final binary developed image. We summarize the forward model, starting from the projections $P(x,\theta)$:

\begin{enumerate}
  \item Quantization of the positive images to 8-bit values to be fed to the blue channel of a digital light projector 
  \item Conversion of 8-bit values to physical intensities (measured in $W/cm^2$ with a calibrated silicon photodiode) 
  \item Conversion of integrated intensities over all angles into a degree of crosslinking (or curing): the fraction of polymer molecules that have crosslinked. Based on the polymerization kinetics described in the next section, it turns out that the rate of curing is proportional to the square root of intensity, with exponential dependence on $\sqrt{I}t$. In the present model however, we assume a simplifying linearization and use accumulated dose as the metric for degree of crosslinking. Better characterization and modeling will be required for a more accurate model depending on particular rate constants. The assumption of accumulated dose irrespective of time history of exposure is more valid if the angle projection is performed simultaneously or if multiple angular iterations are performed in the same exposure.
 \item The final step is the development process, which washes away uncured material. Starting with degree of cure, we use a threshold or gelpoint to model the effect of the developing solution. This converts a continuous degree of cure into a binary function via a thresholding operation. While this leads to a non-convex objective function, the sharp thresholding can help recover sharpness that is lost due to the positivity constraint and sampling issues. 
\end{enumerate}

The goal of the optimization algorithm is to calculate the set of backprojections $P_{opt}(x,\theta)$ that best produces a desired output intensity. We use an optimization procedure based on \textbf{projected gradient descent}. This is guaranteed to converge for a convex objective and convex constraint set. Neither of these is true in our case due to the thresholding function and discrete values for the projector input. However, this heuristic performs quite well for some simple geometries, and reasonably well for more complicated ones. One iteration of the optimization loop to generate $P_{n+1}(x, \theta)$ from $P_{n}(x,\theta)$ given a target image $R(x,y)$ goes as follows (illustrated in Fig. 4):

\begin{enumerate}
\item \textbf{Projection:} Starting with the 8-bit DLP projections $P_{n}(x,\theta)$ generate an unthresholded 2D dose distribution of power $D(x,y)$. The calibration for DLP value to power comes from measurements taken with a calibrated silicon photodiode. We take $D(x,y)$ to be the degree of crosslinking. A future implementation will directly solve the photopolymerization equation integrated over time and space to generate the degree of crosslinking in the forward model. 
\item \textbf{Thresholding:} Depending on the development recipe, convert the degree of crosslinking to the thresholded image at the nth iteration $R_n(x,y)$. In the discussion section we describe variations to this procedure, to prevent the optimization from being sensitive to a single threshold, and instead attempting to penalize absolute errors around the threshold as well. In all cases however, $R_n(x,y)$ lies between 0 and 1, representing material fully washed away as opposed to fully present
\item \textbf{Error:} We determine the error in the image by comparing with the target: $\delta_n(x,y) = R_n(x,y) - R(x,y)$. We transform this into the backprojection domain by performing a radon transform (integral projection) at every angle, followed by ramp filtering. This leads to the projection domain error $\delta^{'}_{n}(x,\theta)$
\item \textbf{Update:} An unconstrained new set of projections is computed as: $R_{n+1}(x,\theta) = R_n(x,\theta) - \delta^{'}_{n}(x,\theta)$. Finally, the computed projection is constrained to positive 8-bit values by first setting negatives to zero, and then quantizing. It can be confirmed that the updated $R_{n+1}(x,\theta)$ is the closest element in the constrained set to the computed unconstrained value and is therefore a \textit{projection}. 
\end{enumerate} 

\begin{figure}[ht]
  \centering
  \includegraphics[width=3.3in]{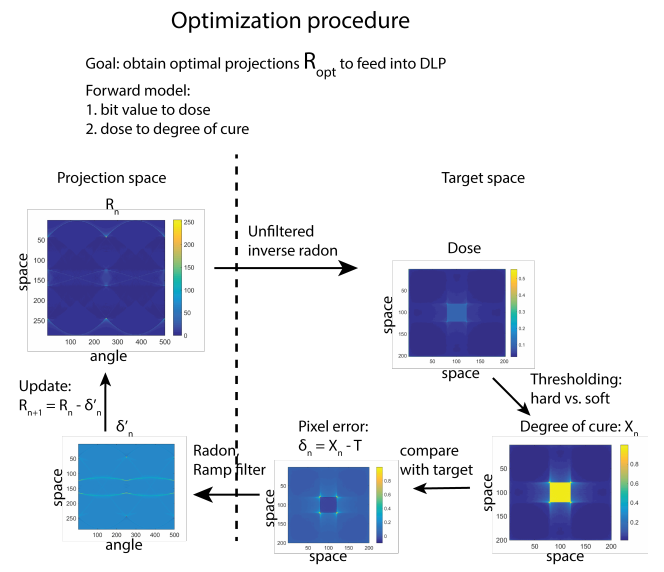}
  \caption{The optimization flow, with example images corresponding to each stage in the algorithm. This illustration is for a softer thresholding case.}
  \label{optimization}
\end{figure}

\subsection{Discussion}

We have tested the optimization routine with some simple target geometries to understand limitations better. Firstly, we were pleasantly surprised to note that the optimization with hard thresholding in the development step leads to perfect reconstruction in some test cases, depending on the number of angular samples. Fig. 5 shows results for three different cases of angular sampling for an image with 500 spatial samples in each dimension. According to the sampling argument in the Fourier domain, a good number of angular samples to have over 180 degrees would be $500\times \pi/2 \sim 785$, corresponding to an angular separation of $0.23^{\circ}$. As the figure shows, we have perfect convergence within 25 iterations when the angular separation is $0.36^{\circ}$, corresponding to 500 angular samples. Coarser angular sampling down to 200 samples over 180 degrees produces reasonable results too. However, 50 angular samples is insufficient.

\begin{figure}[ht]
  \centering
  \includegraphics[width=3.3in]{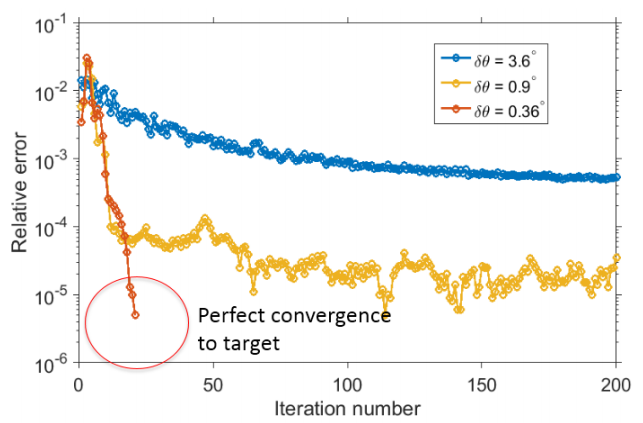}
  \caption{The effect of angular sampling on the image formation. The plotted error is normalized to number of pixels in the 2D image. The target image was a rectangle.}
  \label{angle_sampling}
\end{figure}

The non-linear thresholding property of the resin is important for sharpness in the reconstructed 2D image since this amplifies variations in dosage and leads to sharp boundaries even when the underlying dose distribution is smooth. Another important test of the resin was to investigate the results when the thresholding function was varied. In reality, the thresholding function is somewhat stochastic and results in a binary result depending on a threshold subject to local spatial variations and randomness in the development process. This variability in threshold results in highly varying final cured images if the optimization exploits small intensity variations around the threshold. Therefore, we wanted to test smooth thresholding functions that penalize variations around threshold and make the dose distribution look more like the target image. We tested two ways of doing this:

\begin{itemize}
 \item Two thresholds $x_l$ and $x_u$ could be used to compute a smoother thresholding function. In case the dose falls below $x_l$ or above $x_u$, then we assign 0 and 1 respectively. For intermediate values, we assign a linearly interpolated value depending on the exact dose. For smooth thresholding functions, we find that the relative error does not go to 0 and increases for larger $|x_u - x_l|$. However, the dosage reproduces the original target better.
\item A sigmoid thresholding function of the form $1/(1+e^{-\frac{x-\mu}{\sigma}})$ could be used for a continuous variation of thresholding parameters. 
\end{itemize}

\begin{figure}[ht]
  \centering
  \includegraphics[width=3.3in]{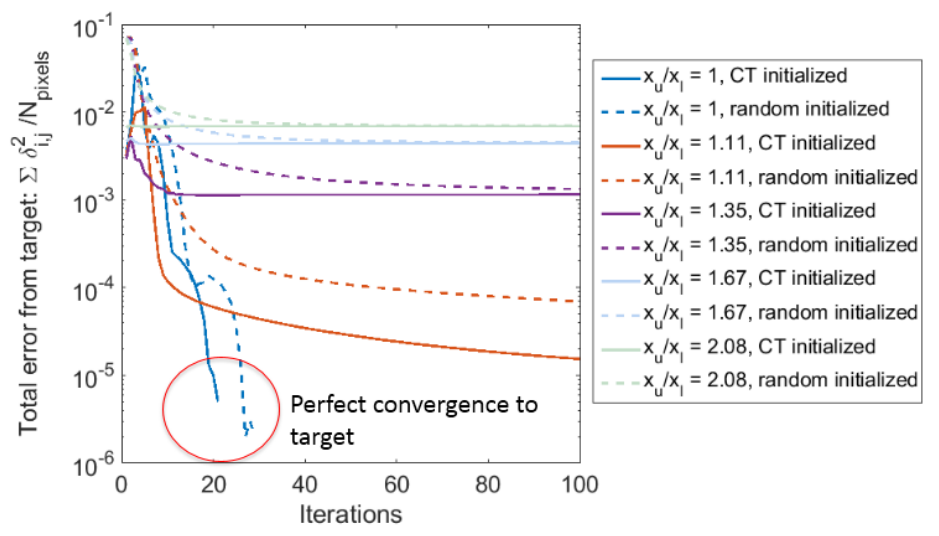}
  \caption{Double threshold convergence: The effect of varying the thresholds, with a sharp threshold of $x_u/x_l = 1$ leading to perfect convergence whereas a smooth threshold showing some error due to the non-linearity being less effective at sharpening the image. We also attempted each case for two different initial conditions: random and CT-based (non-negativized projections). The CT case converges slightly faster.}
  \label{doublethresh}
\end{figure}

The effect of smooth thresholding is particularly obvious in the sigmoid thresholding case. If we optimize using a sharp sigmoid (small $\sigma$), then convergence is slower, the nonlinearity better recovers the final output - but the unthresholded image has lower contrast and relies on the properties of the resin more. However, in the case of the smooth sigmoid, convergence is faster - leading to an image with greater contrast. In our final tests, we decided to use the more robust soft thresholding method. However, in case we decide to move to a high contrast photoresist (such as SU-8) with a more reproducible and sensitive response, we may be able to benefit from the accurate reproduction of the target that the nonlinearity provides.

\begin{figure}[ht]
  \centering
  \includegraphics[width=3.3in]{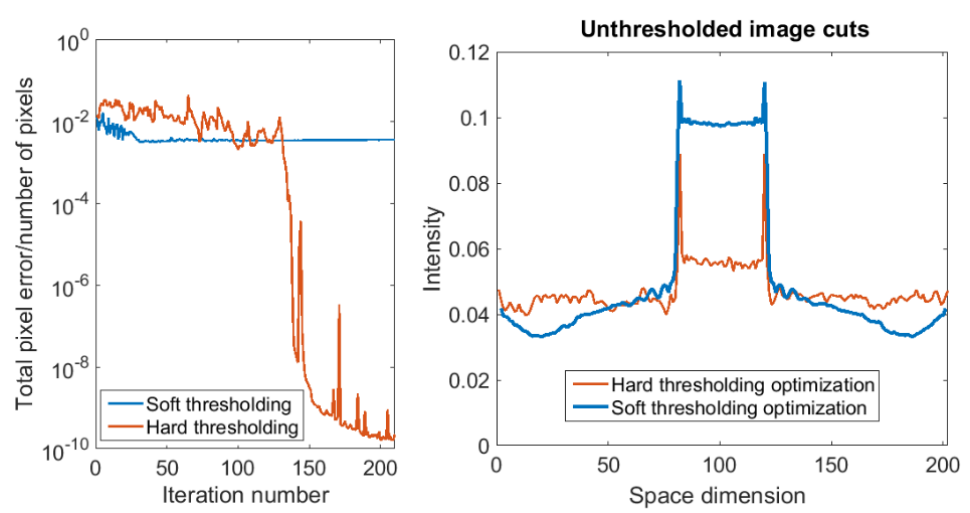}
  \caption{Sigmoid thresholding: Left - comparison of convergence for the soft vs. hard thresholding case. Right - 1D cuts through the unthresholded dose distribution for the soft vs. hard thresholding. The target image was a rectangle.}
  \label{sigmoid}
\end{figure}

\section{Chemical Resin Response}

\subsection{Resin Formulation}

One important aspect in the design of a 3D tomographic lithography system is to understand and exploit the polymer photocrosslinking chemistry. In this work, photcurable acrylate polymers are used with free radical photoinitiators to induce crosslinking in the base polymers. Multiple resin formuations were used. In the 2D algorithm validation, the prepolymer Triethylene glycol dimethacrylate (TEGDMA) was used, with 1.0 wt\% photoinitiator Camphorquinone (CQ) and 0.5 wt\% co-initiator ethyl 4-(dimethylamino)benzoate (EDAB). Later, for 3D printing validation, a mixture cosisting of 75 wt\% Bisphenol A glycerolate (1 glycerol/phenol) diacrylate with 25 wt\% Poly(ethylene glycol) Diacrylate (PEGDA) 250Da and 0.4 wt\% photoinitiator Irgacure 819 is used.  All chemicals were obtained from Sigma Aldrich.

\subsection{Resin Calibration}
\label{Resin Calibration}

Both resins used in this work are activated by absorption of blue light from a DLP projector the photoinitiator. This triggers a reaction which generates free radicals in the (meth)acrylate end groups of the prepolymer which then form covalent crosslinks between polymer chains. As the exposure dose is increased, more radicals are generated and more crosslinks form. As the local density of crosslinks, or degree of cure, increases, the polymer material transitions from a liquid to a solid and increases in stiffness. The liquid-solid transition can be characterized by the gel point, a threshold in the degree of cure above which the material's storage modulus exceeds its loss modulus. Near this point there exists a threshold in the degree of cure below which material will be washed away in a development step and above which the material remains. The development steps used in this work were rinses in isopropyl alcohol (IPA) or Acetone.

The classical equation for free radical photopolymerization in bulk, Equation \ref{ClassicalRp} \cite{Andrz}, can be used initially as a model for the resin response where the parameters $k_p$ and $k_t$ are, respectively, the rate parameters for propagation and termination, $\left[M\right]$ is the initial prepolymer concentratrion, $\phi$ is the quantum yield, and $I_a$ is the absorbed intensity of the source.
 
 \begin{equation}\label{ClassicalRp}
R_p = \frac{k_p}{k_t^{1/2}}\left[M\right]\left(\phi I_a\right)^{1/2}
\end {equation}

As the reaction proceeds and crosslinks are formed, the prepolymer concentration decreases at the propagation rate $R_p$ through Equation \ref{RpM}

 \begin{equation}\label{RpM}
R_p = -\frac{d\left[M\right]}{dt}
\end {equation}

Defining the degree of crosslinking ($DOC$) as the ratio crosslinks formed to available reaction sites, we can write the $DOC$ in terms of the initial and current prepolymer concentrations $\left[M\right]$ and $\left[M\right]_0$

 \begin{equation}
DOC = 1 - \frac{\left[M\right]}{\left[M\right]_0}
\end {equation}

For constant intensity we can solve for the degree of cure vs. time.

 \begin{equation}\label{DOCvstime}
DOC = 1 - exp\left(-\frac{k_p}{k_t^{1/2}}\left(\phi I_a\right)^{1/2}*t\right)
\end {equation}

As can be seen from Equation \ref{DOCvstime}, the degree of cure is a function of both intensity and exposure time and not simply exposure dose. Additionally, it is know that free radical inhibitors present within the resin are highly reactive with radicals and will tend to quench radicals before they can initiate a cross-linking reaction. This happens until the inhibitor concentration is significanltly reduced to the point where the initiation reaction between the initiator and acrylate group can compete with the inhibition reaction. In effect, enough radicals must be generated in a particular location to first deplete the inhibitor concetration before crosslinking can begin. While only very small concentrations of inhibitor molecules are added to the liquid prepolymers, diatomic oxygen also acts as a free radical inhibitor. It is believed that a small concentration of oxygen, on the order of ~ 0.001M is present with in the resin from the start. The local oxygen concentration must be overcome before curing begins. In low viscosity resins, where the diffusivity of oxygen is higher, this can lead to proximity effects.

\begin{figure}[ht]
  \centering
  \includegraphics[width=3.0in]{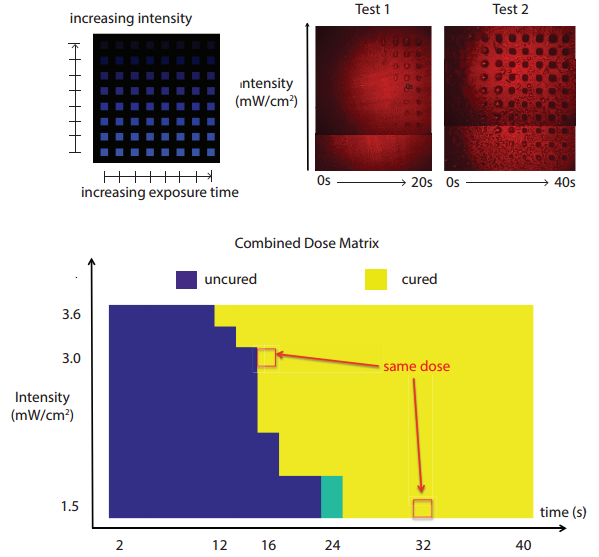}
  \caption{Time-Intensity dose matrix used to calibrate resin response. Each square receives a different combination of intensity and exposure duration. Results from two tests are combined graphically. Results show that curing behavior is a function of intensity and time, not simply expsoure dose}
  \label{DoseMatrix}
\end{figure}

In order to probe this chemical relation a test was designed where a thin layer of the TEGDMA resin was exposed to an array of square features at varying intensity values and exposure times. After development with IPA, a binary map of cured vs. uncured features was generated and is plotted in Fig. \ref{DoseMatrix}. The results confirm that the curing behavior is a function of both intensity and time and not simply net exposure dose $\int I_a dt$. Further investigation and modeling into the thresholding behavior of the polymer with respect to development as well as the evaluation and measurement of rate constants $k_p$ and $k_t$ is ongoing.

\section{CAL Algorithm Validation}

\subsection{Algorithm Test System}

\begin{figure}[ht]
  \centering
  \includegraphics[width=3.3in]{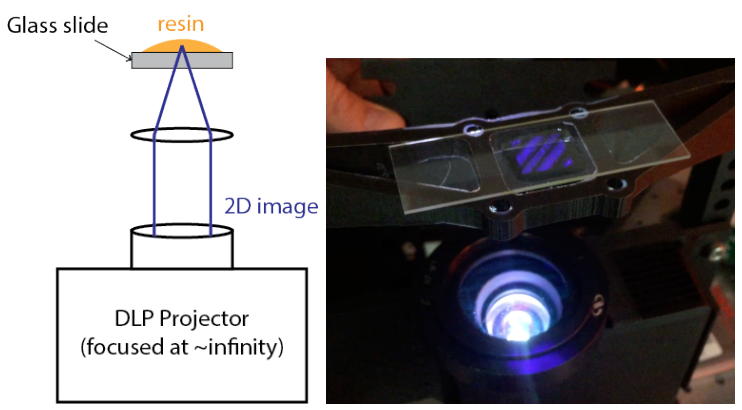}
  \caption{2D CAL printing system used for initial tests on algorithmioc design of projections}
  \label{slice_to_2Dproj}
\end{figure}

\begin{figure}[ht]
  \centering
  \includegraphics[width=3in]{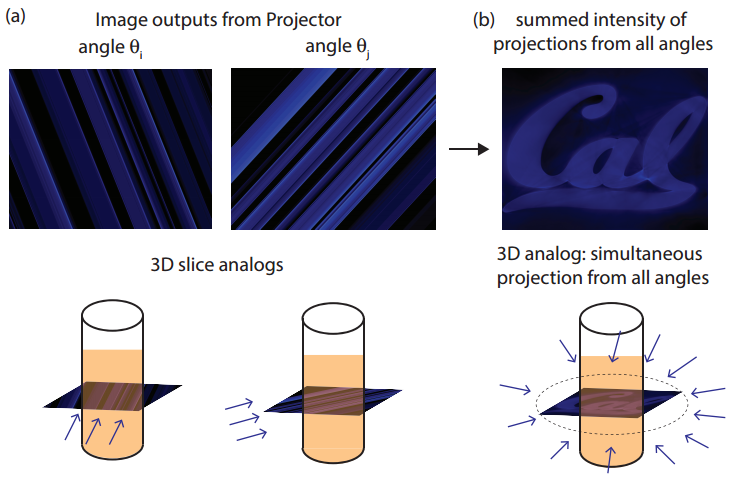}
  \caption{Representation of the analogy between the 2D and 3D CAL prototype systems}
  \label{HardwareSimulation_crossSection}
\end{figure}

Before attempting 3D CAL printing, an experimental apparatus was constructed to directly test the projection design algorithm. This was done to perform an initial validation of the algorithm before accounting for 3D effects such as optical attenuation and gravity. For this apparatus, a DLP projector was used as the optical source to generate the projections. The system was configured as shown in Fig. \ref{slice_to_2Dproj} such that a 2D image was incident on a thin layer of resin. The video output from the projector was set to match the intensity distribution which would be incident upon a cross-section of the 3D volume in a sequential illumination CAL system as illustrated in Fig. \ref{HardwareSimulation_crossSection}. Thus, each video frame corresponded to a 1D projection propagated in a single direction dictated by angle of illumination from the algorithm.

Using this prototype CAL system, the algorithm was tested for its ability to print various geometries. Projections from 500 evenly spaced angles about $360^o$ were computed and used to generate the video frames such as the ones illustrated in Fig. \ref{HardwareSimulation_crossSection}a. The frame rate of the projected movie was set to simulate an angular rotation of $25^o/s$ to match the max rotational speed of the rotating stage used in the 3D CAL prototype described in Section \ref{sec: 3D setup}.

\begin{figure}[ht]
  \centering
  \includegraphics[width=3in]{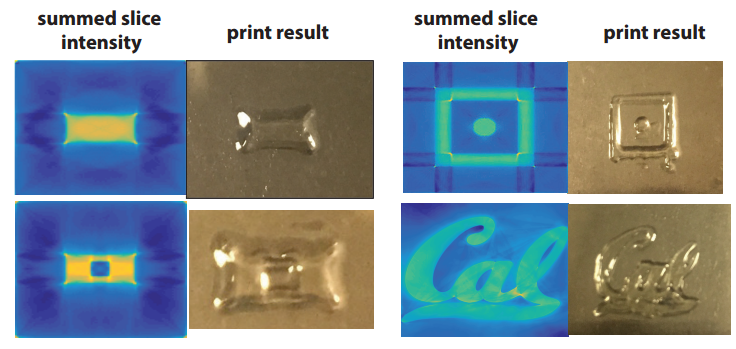}
  \caption{Results from hardware simulation}
  \label{Results}
\end{figure}

\subsection{Algorithm Test Results}

\begin{figure}[ht]
  \centering
  \includegraphics[width=3in]{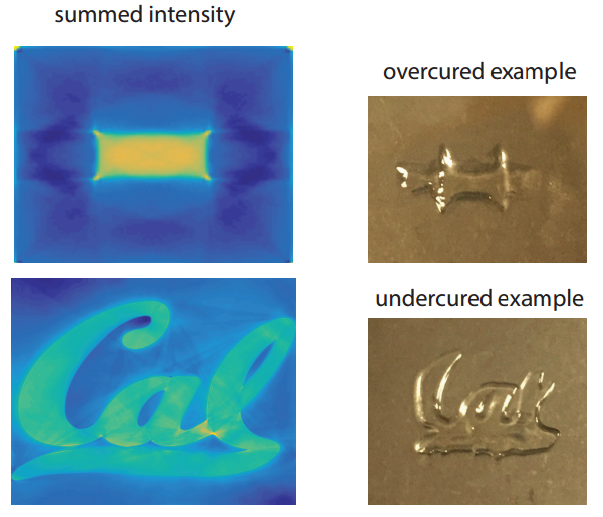}
  \caption{Results from hardware simulation}
  \label{BadResult}
\end{figure}

The 2D CAL prototype yielded promising results to validate the CT algorithm. These results are given in Fig. \ref{Results}. It should be restated that when performing this hardware simulation it is necessary to monitor and control the total exposure dose, in this case the number simulated rotations of the vial, so that the thresholding behavior is properly taken advantage of. For illustration, examples of cases where the exposure was either too low or too high are given in Fig. \ref{BadResult}. If the exposure is too low, regions within the target image which have lower dose do not develop. If the exposure is too high, regions outside of the target image which have relatively high dose begin to develop.

\section{CAL 3D Printing}

\subsection{3D CAL Printing System}
\label{sec: 3D setup}
\begin{figure}[ht]
  \centering
  \includegraphics[width=3.0in]{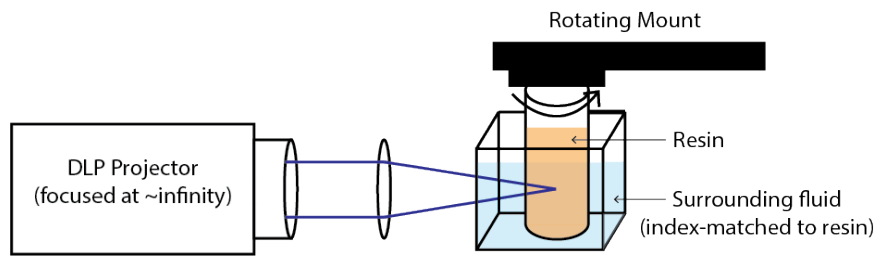}
  \caption{CAL 3D printing system schematic. DLP projector projects 2D images frame-by-frame in the form of a video while the resin volume rotates to change the relative angle between resin-centric coordinate system and the projected image}
  \label{3DTomogProto_schematic}
\end{figure}

In order validate the ability of CAL to print 3D parts, a second prototype system was constructed. This system is depicted in Fig. \ref{3DTomogProto_schematic}. The major hardware design decision in building a CAL system comes from the delivery of the optical image to the resin. Ideally, many images illuminate the target volume simultaneously from many angles about the central axis. A design concept for how to achieve this from hardware is presented later in this paper and drawn in Fig. \ref{FutureSystems}. For ease in generating a prototype, however, the image delivery mechanism was constructed from the DLP projector used in the 2D prototype and a rotating cylindrical vial containing a photopolymer resin. In the 3D prototype system, as the cylindrical target volume rotates, the image output from the projector is switched. The output comes in the form a video where each frame is the image output from the algorithm for a particular angle. The incoming angle is set by the rotation of the vial. To avoid cylindrical lensing effects which distort the incident image, a rectangular box filled with a fluid of higher refractive index was used. Ideally, the fluid is index-matched to the photopolymer resin to negate the lensing effects. In this work, the index matching fluid used was the same prepolymer composition used in the resin but without the photoinitiator added.

\subsection{3D effects}

When moving from the 2D to the 3D prototype, additional physical effects become relevant that make 3D CAL printing less straightforward. For one, when the incident illumination now hits a much thicker, absorptive resin, the intensity at the center and back end of the vial are less than at the front. Additionally, in the 3D system, as the part begins to cure and the density increases there are additonal physical forces (gravity and centripetal) on the curing piece. Finally, as the resin begins to cure, the refractive index changes and scattering effects arise. We have addressed each of these effects to allow for succesful 3D CAL printing.

First, to address the effect of forces exerted on the curing part, a new resin was used which had much higher viscosity. This polymer used in this resin comprised of a mix of 75 wt\% Bisphenol A glycerolate (1 glycerol/phenol) diacrylate and 25 wt\% PEGDA 250 Da. The viscosity was mesured using cone and plate rheometry to be ~4000 cP and independent of shear rate.

\begin{figure}[ht]
  \centering
  \includegraphics[width=3.0in]{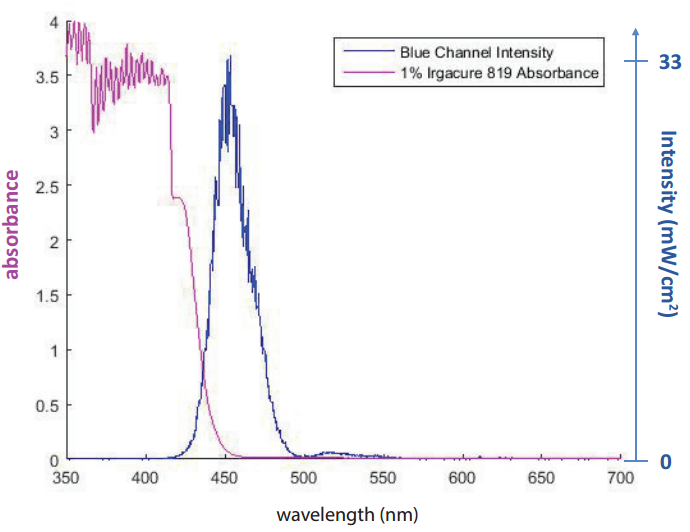}
  \caption{Absorbance curve of 3D CAL resin plotted on the same axes as the spectral output of the DLP source}
  \label{ResinAbsorbance}
\end{figure}

Second, to address the absorption issue, a photoinitiator with lower absorption in the illuminating spectral regime, Irgacure 819 was used. It was also mixed at a lower concentration, 0.4 wt\%. The absorbance of the new resin was measured using a UV-Vis spectrophotometer. In Fig. \ref{ResinAbsorbance}, the measured absorbance is plotted against the measured spectral output of the DLP projector's blue channel.

Finally, because of the oxygen inhibition period described in section \ref{Resin Calibration}, the resin is able to record exposure, through local depletion of oxygen, before curing begins. Thus, when scattering effects begin to occur the geometry has already been defined through local inhibitor depletion.

\subsection{3D Results}

\begin{figure}[ht]
  \centering
 \includegraphics[width=3.3in]{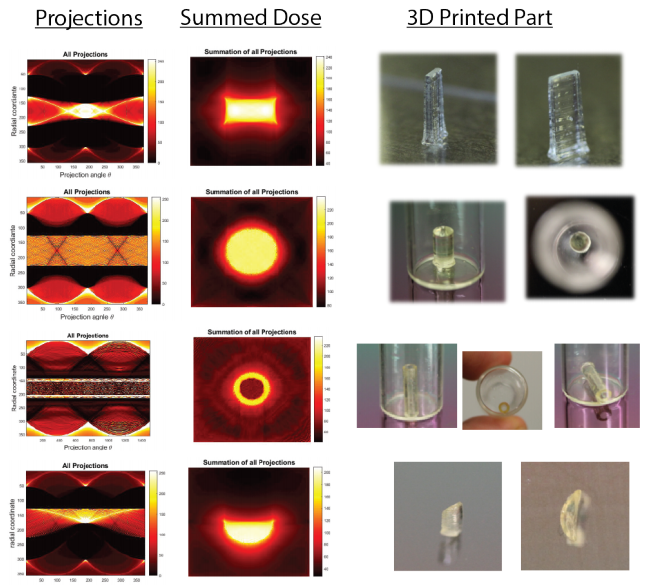}
  \caption{3D CAL printed results with constant cross-section in z. Column 1: 1D projections vs angle for each z slice. Column 2: summed intensity distribution from all projections in each z slice. Column 2. Images of 3D printed results}
  \label{3DResults}
\end{figure}

Using the 3D CAL prototype system, succesful 3D prints of various geometries were achieved. These results are displayed in Fig. \ref{3DResults}. Four parts were printed. Each has a constant cross-section in the verical (z) dimension and was thus printed from a stacking of 1D projections. The first geometry, a rectangular prism, demonstrates some ability of CAL to print geometries with sharp corners. Another geometry, with a semicircle cross-section, demonstrates the ability to simulaneously print curved surfaces, flat surfaces, and corners. Most importantly, successful printing of a hollow cylinder demonstrates the ability to print voids within a solid structure. The next step that will be pursued on the 3D CAL prototype system is the printing of fully three dimensional parts, those which have varying cross-section in z.

\section{Discussion/Future Work}

\begin{figure}[ht]
  \centering
  \includegraphics[width= 3.3in]{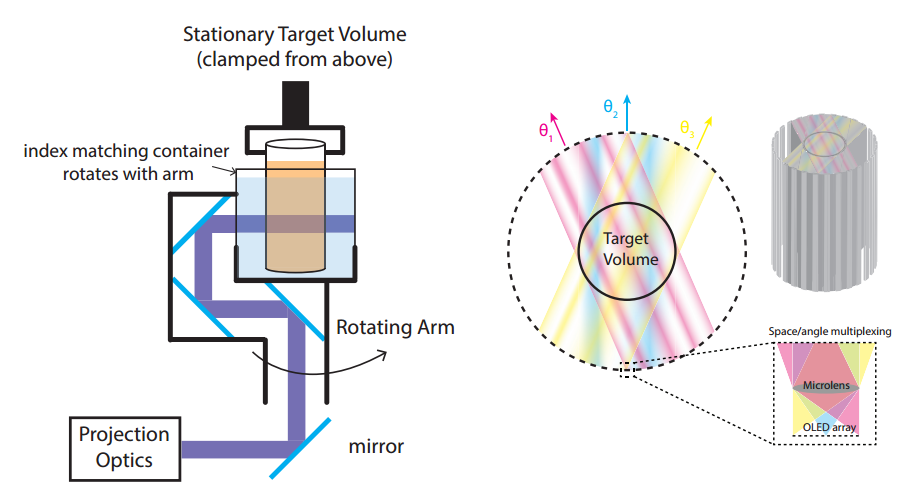}
  \caption{Potential future systems designs for CT lithography. Left: Stationary target volume with rotating optics. Optics can be rotated much faster than resin as fluid motion in resin is no longer a factor. Right: light field projection system for true single-shot CT lithography. Images from all angles are projected simultaneously}
  \label{FutureSystems}
\end{figure}

The present work demonstrates a careful design and initial prototyping of an additive manufacturing system which can promote a new paradigm in the way such systems are designed and operate. It fosters a shift from the conventional approach of building 3D parts layer-by-layer toward an approach which is designed to address the entire volume of the 3D geometry. This type of shift can present advantages in the speed of manufacture of parts as well as in the ability to print overhanging geometries without the need for mechanical support material. To further fuel this shift, there are some logical amendments to the prototyped system that will be implemented in future work. First, some improvements can be made by the design of a system where the resin volume is static and the optics rotate around it. The advantage of this system design is that the optics can be rotated at a much higher speed than the volume. This arises from the constraint of fluid motion in the resin volume at higher rotation speeds, which. The only constraints on the speed with which the optics can be rotated are motor rotation speed limits and the minimum frame rate of the optical system. Neither limit has been nearly approached in the current prototype. A design concept for this work can be seen in Fig. \ref{FutureSystems}. Additionally, to achieve true single-shot 3D lithography, we propose to design a system such as the one depicted in Fig. \ref{FutureSystems} (right panel) which could produce projections from many angles simultaneously. We also plan to continue using the 2D CAL prototype as a test platform for a better understanding of resolution limitations and the reproducibility of the printing recipe. In the dose matrix test of Fig. \ref{DoseMatrix}, we already observe that the boundaries of the square projected patterns are rounded. We speculate that this could be due to oxygen diffusion that spatially blurs the solidification threshold. Additionally, the imaging optics used has aberrations, especially near the boundary of the developed dose matrix test. Finally, the pixel size of the projector leads to a limitation on the spatial resolution of this system. These effects will be explored in the future, along with the development of a more reproducible experimental procedure. 

To conclude, we would like to describe one possible embodiment of a single-shot 3D Lithography system. This is inspired by the Lytro plenoptic camera that images an angular as well as spatially resolved ray space \cite{Lightfield}. Prior work on near-eye lightfield displays also implements a related system \cite{NearEye}. Here, in order to simultaneously illuminate all angular pixels as well as spatial pixels, we would like to use a microlens array, with subpixels of LEDs, placed at the focal plane under each microlens. The microlens array could wrap around the target volume in a concentric manner as shown in Fig. \ref{FutureSystems}, right panel. The inset shows a single microlens with multiple LED subpixels. In the tomographic construction procedure, we would like to have both a high spatial and angular resolution. If we consider the illustrated geometry, this allows us to enumerate a tradeoff between angular and spatial resolution. For concrete numbers, let us consider a 10 cm target volume radius and 30 cm outer radius. The spatial pixel shown in the inset is required to project the set of angles from $\theta_1$ to $\theta_3$ given by the over all geometry of the system. In this example, $-\theta_1 = \theta_3 = 18.5^{\circ}$. We have previously discussed a sampling consideration which suggests that the number of angular samples over $180^{\circ}$ should be a factor of $\pi/2$ higher than the number of spatial samples. In this example, if we were to use 500 spatial samples in each transverse dimension, this would lead to 785 angular samples. The angular spacing is then $0.23^{\circ}$, which leads to 160 angular subpixels under one spatial pixel. With 500 spatial samples spread across a target region of size 10 cm, this leads to a $200 \mu m$ spatial pixel size, with the LED size given by $1.25 \mu m$. The LED size could potentially be scaled below the diffraction limit, but typical individually addressed AMOLED arrays tend to be around 10 microns in size. 

To generalize slightly and explicitly consider how this implementation will scale, we consider a target radius of $r$, outer radius of $R$, minimum subpixel size of $\lambda$ and $N$ the number of spatial pixels in each dimension. Then, the number of angular samples is $\pi N/2$, leading to an angular sample every $2/N$ radians. Thus, the number of subpixels under one microlens will be $N tan^{-1}(r/R)$, with the microlens size being $r/N$. This leads to the subpixel size $\lambda = r/(N^2 tan^{-1}(r/R))$. Therefore, if we set the spatial resolution, target volume size and minimum subpixel size, we can calculate the required number of samples and outer radius. For instance, in order to print a 5 cm target radius with 100 $\mu m$ resolution using 10 $\mu m$ subpixels would require 1000 spatial samples and an outer radius of 10 m. The impractically large outer radius helps to achieve a small angular resolution using a limited subpixel size. This suggests that reducing the number of angular samples while maintaining spatial resolution is going to be an important problem to address for this implementation. The non-linear thresholding properties of a sensitive resin may well be one solution to help achieve accurate reconstruction with sparse angular sampling.

\section*{Acknowledgements}

This work was performed under the auspices of the U.S. Department of Energy of Lawrence Livermore National Laboratory under Contract DE-AC52-07NA27344. This work was performed under LDRD funding 14-SI-004, and 17-ERD-116. This document has been released under IM review \# LLNL-JRNL-731365. We would like to thank both the Design for Nanomanufacturing group at the University of California, Berkeley and the Center for Engineered Materials and Manufacturing at Lawrence Livermore National Laboratory. We would also like to acknowledge useful discussions with Prof. Laura Waller, Jingzhao Zhang and Prof. Ren Ng at Berkeley and with Allison Browar, James Oakdale, and Ryan Hensleigh at LLNL.

\bibliography{ArXiv_bib}

\end{document}